
\headline={\ifnum\pageno=1\firstheadline\else
\ifodd\pageno\rightheadline \else\leftheadline\fi\fi}
\def\firstheadline{\hfil}
\def\rightheadline{\hfil}
\def\leftheadline{\hfil}
        \footline={\ifnum\pageno=1\firstfootline\else\otherfootline\fi}
\def\firstfootline{\rm\hss\folio\hss}
\def\otherfootline{\hfil}

\font\twelvebf=cmbx10 scaled\magstep 1
\font\twelverm=cmr10 scaled\magstep 1
\font\twelveit=cmti10 scaled\magstep 1

\font\tenbf=cmbx10
\font\tenrm=cmr10
\font\tenit=cmti10

\parindent=1.5pc
\hsize=6.0truein
\vsize=8.5truein
\nopagenumbers

\def\to{\rightarrow}
\def\bar{\overline}
\def\CO{{\cal O}}

\def\NPB#1{{\twelveit Nucl. Phys.}       {\twelvebf B#1}}
\def\PLB#1{{\twelveit Phys. Lett.}       {\twelvebf B#1}}
\def\PRD#1{{\twelveit Phys. Rev.}        {\twelvebf D#1}}
\def\RMP#1{{\twelveit Rev. Mod. Phys.}   {\twelvebf  #1}}
\def\ZPC#1{{\twelveit Zeit. Phys.}       {\twelvebf C#1}}

\line{\hfill IISc-CTS-3/94}
\line{\hfill hep-ph/9405213}
\bigskip

\centerline{\tenbf FROM KONDO MODEL AND STRONG COUPLING LATTICE QCD}
\baselineskip=16pt
\centerline{\tenbf TO THE ISGUR-WISE FUNCTION\footnote{$^\dagger$}{%
\rm Invited talk presented at the International Colloquium on
Modern Quantum Field Theory II, TIFR, Bombay, January 5-11, 1994.}}

\vglue 0.8cm
\centerline{\tenrm APOORVA PATEL\footnote{$^*$}{%
E-mail: adpatel@cts.iisc.ernet.in}}
\baselineskip=13pt
\centerline{\tenit CTS and SERC, Indian Institute of Science}
\baselineskip=12pt
\centerline{\tenit Bangalore-560012, India}

\vglue 0.8cm
\centerline{\tenrm ABSTRACT}
\vglue 0.3cm
{\rightskip=3pc
 \leftskip=3pc
 \tenrm\baselineskip=12pt\noindent
Isgur-Wise functions parametrise the leading behaviour of weak decay
form factors of mesons and baryons containing a single heavy quark.
The form factors for the quark mass operator are calculated in strong
coupling lattice QCD, and Isgur-Wise functions extracted from them.
Based on renormalisation group invariance of the operators involved,
it is argued that the Isgur-Wise functions would be the same in the
weak coupling continuum theory.
\vglue 0.6cm}

\vfil
\twelverm
\baselineskip=14pt

Quantum Chromodynamics with heavy quarks possesses spin-flavour
symmetries that become exact as the quark masses go to infinity.
These symmetries give rise to relations amongst various matrix
elements and form factors of hadrons containing heavy quarks$^1$.
Such relations based on symmetry properties alone are genuine
predictions of QCD, and do not suffer from the uncertainties of
phenomenological models of hadrons. Of course, the leading order
relations (i.e. those valid in the $M\to\infty$ limit) have to
be corrected for symmetry breaking effects in order to connect
them to properties of physical hadrons containing heavy quarks.
These corrections arise from unequal quark masses and from terms
suppressed by powers of $1/M$, and are in the range of $10-20\%$
for many instances involving $b$ and $c$ quarks. Thus extracting
the leading behaviour using heavy quark symmetries, and then
estimating the corrections using some phenomenological model,
is a practical solution to cut down our ignorance in dealing
with QCD and to make useful predictions for fitting experimental
results concerning hadrons containing heavy quarks$^2$.

Of particular phenomenological interest are the weak decay form
factors of hadrons containing a single heavy quark. Together with
the experimentally observerd weak decay matrix elements, these form
factors determine various elements of the Cabbibo-Kobayashi-Maskawa
quark mixing matrix. For example, a precise determination of the
element $V_{cb}$ is possible provided we know the form factors of
vector and axial currents between $B$ and $D$ mesons and $B \to D$
semileptonic decay rates. The approach outlined above predicts that
in the leading order all such form factors can be reduced to two
unknown functions, one for mesons ($\xi$) and another for baryons
($\zeta$). These two we refer to as the Isgur-Wise functions.

The symmetry properties of hadrons containing a single heavy quark
are best expressed in the language of an effective field theory.
In this formalism, factors of heavy quark mass are explicitly taken
out by appropriately scaling all variables and the hadronic states
are characterised by their four-velocities. In the $M\to\infty$ limit,
the momentum carried by the QCD degrees of freedom (gluons and light
quarks) that interact with the heavy quark is too small to alter
$v_\mu$, and so $v_\mu$ becomes a conserved quantum number. Of course,
the large momentum transfer involved in a weak decay can change $v_\mu$
by a finite amount. $\xi$ and $\zeta$ are thus functions of the only
Lorentz invariant combination available, $v\cdot v'$, where $v$ and
$v'$ are the heavy hadron velocities in the initial and final states
respectively. (Note that $v\cdot v'$ is nothing but the Lorentz
contraction factor $\gamma$ of special relativity, while $v^2=1=v'^2$.)

In the rest frame of the heavy hadrons, the heavy quark just sits
at one point, say the origin, while the light QCD degrees of freedom
buzz around it. The situation corresponds to embedding a static colour
impurity in the QCD vacuum. This picture provides explicit information
about the functions $\xi$ and $\zeta$ in two specific geometries,
$v=v'$ and $v=-v'$. When $v=v'$, the heavy quark may decay at some
instance of time, but the surrounding light QCD degrees of freedom
do not feel any change at all. The flavour independence of QCD thus
allows a convenient choice for the absolute normalisation of the
Isgur-Wise functions, $\xi(v\cdot v'=1) = 1 = \zeta(v\cdot v'=1)$.
When $v=-v'$, the actual physical process is heavy quark pair creation
or annihilation. So we expect the Isgur-Wise functions to be singular
at this kinematic point. Analytically continuing the Isgur-Wise
functions to the full complex $v\cdot v'$ plane, a minimal scenario
for their singularity structure is to have a pole at $v\cdot v' = -1$,
followed by a branch cut for $v\cdot v' < -1$, and no other singularities
in the rest of the complex plane. The physical situations of course
correspond to $|v\cdot v'| \ge 1$ along the real axis. It is important
to note that both these constraints on the Isgur-Wise functions remain
unaffected by QCD renormalisation effects. An easy way to see this is
to go to the temporal gauge $A_0=0$, in which the static heavy quark
loses all its QCD interactions and behaves like a free fermion.

To go beyond these constraints and explicitly determine the Isgur-Wise
functions requires knowledge of the dynamical behaviour of QCD. Formally
the set up has several similarities to the Kondo problem of condensed
matter physics, i.e. a static magnetic impurity embedded in a free
electron conduction band. To name a few:
\item{$\bullet$}{Both problems have a static impurity interacting with
its polarisable surroundings. Kondo problem has spin interaction, while
heavy quark QCD has colour interaction. Only $s-$wave configurations
feel the interaction in the leading order, and they are the cases of
phenomenological importance.}
\item{$\bullet$}{The interaction is weak at short distances, but becomes
strong at long distances completely screening the impurity. The change
from asymptotic freedom to confinement is smooth without any intervening
phase transition, and one can talk about renormalisation group (RG) flows
leading from the weak coupling fixed point to the strong coupling one.}
\item{$\bullet$}{The unstable weak coupling fixed point is at zero coupling,
with an essential singularity and a logarithmically running coupling
governing the scaling behaviour. The stable strong coupling fixed point
is at infinite coupling, and is of a trivial nature. There is no inherent
mass scale in either problem (the light quark masses can be set to zero
in case of QCD).}
\item{$\bullet$}{The scale characterising the cross-over from weak to
strong coupling is dynamically generated. This dimensional transmutation
can be described as the irrelevant operators determining the absolute
scale of the theory. The Kondo temperature and $\Lambda_{QCD}$ are
defined in terms of the couplings of an ultraviolet regulated theory.}

\noindent
There are substantial differences between the two problems too:
\item{$\circ$}{In the Kondo problem, the electrons interacting with the
magnetic impurity are free, with no interactions amongst themselves
in the absence of the impurity. The electron energy spectrum extends
continuously all the way down to $E=0$. On the other hand, QCD is a
non-trivial theory without any known solution even in the absence of
a static impurity. Its states are discrete with a non-zero mass gap
(ignoring the Goldstone bosons).}
\item{$\circ$}{In the Kondo problem, the objects of study are the
modifications of electronic properties caused by the static impurity.
These can be studied in $s-$wave configurations, while reducing the
$3+1$ dimensional problem to a $1+1$ dimensional one. For heavy quarks
in QCD, the objects of prime importance are not the changes in QCD
excitation spectrum due to the heavy quarks, but the behaviour of the
static colour sources themselves. This behaviour is too simple in the
$v=\pm v'$ geometries, as discussed above. Dynamical features of
interest involve $v\ne\pm v'$, and these geometries cannot be
reduced from $3+1$ dimensions to $1+1$ dimensions.\footnote{$^{\S}$}{%
It is possible to argue that arbitrary functions of $v\cdot v'$ can be
Taylor expanded around the point $v\cdot v'=1$, and the derivatives can
be evaluated in a purely $s-$wave geometry. We then have to calculate
matrix elements of the original operators plus their higher dimensional
descendents defined by insertions of covariant derivatives in the
original operators (such towers of operators are familiar objects in
the operator product expansion and conformal field theories).}}

The modern solution to the Kondo problem is described in the language
of conformal field theory. The differences listed above, however, make
it unlikely that the problem of heavy quarks in QCD can be solved by
the same techniques. On the contrary, the first solution to the Kondo
problem was provided by Wilson in the framework of the renormalisation
group$^{3,4}$, and that is an approach which can definitely be applied
to QCD. We thus turn our attention to the logic used by Wilson in
solving the Kondo problem.

The weak coupling and strong coupling expansions for the quantities of
interest were known in the case of the Kondo problem. Wilson used
RG to interpolate between the two, fixing the ratio of the dimensionful
scale parameters appearing in the two solutions. This needed numerical
RG integration keeping track of the low lying states of the problem.
The desired electronic properties could then be evaluated numerically.
The problem of heavy quark QCD is simpler in a sense, because it has
operators which are RG invariant. Among these operators are partially
conserverd vector and axial currents and their divergences---precisely
the objects which are used to define the Isgur-Wise functions. The task
is to implement the renormalisation group for QCD in such a way that
the desired functions appear in RG invariant matrix elements, evaluate
the matrix elements in the strong coupling limit, and then just read off
the desired functions from the results by separating out the appropriate
proportionality factors.

This procedure can be followed in the lattice formulation of QCD$^5$. The
formulation is non-perturbative, yet at any non-zero value of the lattice
spacing one can define a partially conserved vector current$^6$. The RG
invariant divergence of the vector current is just the difference of the
quark mass operators. The partially conserved vector current is a non-local
object on the lattice and its precise form depends on the details of the
lattice action (it is a $1-$link long operator in the simplest lattice
discretisations), but its divergence is a local operator whose structure
can be chosen independent of the details of the lattice action.

For concreteness, let us consider the vector current form factor for
$B\to D$ semileptonic decay and its limiting behaviour as the $b$ and
$c$ quark masses go to infinity.
$$\eqalignno{
<D(p') | \bar{c} \gamma_\mu b | \bar{B}(p)>
  &~=~ (p+p')_\mu f_+(q^2) ~+~ q_\mu f_-(q^2) \cr
  &{\buildrel M\to\infty \over \longrightarrow}~
  \sqrt{M_B M_D} ~(v+v')_\mu~ \xi(v\cdot v') ~~. &(1) \cr}
$$
Here $q=p-p'$ is the momentum transfer, and $v\cdot v' = (M_B^2 + M_D^2
- q^2)/2M_B M_D$. The divergence of the above expression gives
$$\eqalignno{
{m_b - m_c \over M_B - M_D} <D(p') | \bar{c} 1 b | \bar{B}(p)>
  &~=~ (M_B + M_D) f_+(q^2) ~+~ {q^2 \over M_B - M_D} f_-(q^2) \cr
  &{\buildrel M\to\infty \over \longrightarrow}~
  \sqrt{M_B M_D} ~(1+v\cdot v')~ \xi(v\cdot v') ~~. &(2) \cr}
$$
Note that the operators appearing in the matrix elements on the left
hand side of these equations are composed of the QCD fields, and not
the rescaled fields of the heavy quark effective theory. This feature
is essential to keep the RG evolution of the matrix elements simple,
e.g. both sides of the equations are RG invariant and $\xi$ extracted
from the above form factors does not have anomalous scale dependence.

These results are still in the continuum language. Following Bjorken,
an upper bound on $\xi'(1)$ is obtained by equating the inclusive sum
of probabilities for decays into hadronic states to the probability
for free quark transition. The matrix elements appearing in Eq.~(1)
give$^{7,8}$ $\xi'(1) \le -1/4$, while those appeaing in Eq.~(2) yield
a stronger constraint $\xi'(1) \le -1/2$. The improvement is due to the
fact that the right hand side has a kinematic factor of $(1+v.v')$ in
the divergence equation instead of $(v+v')_\mu$ in the vector current case.

Analogous expressions can be written down for form factors of baryons
containing a single heavy quark. For instance, the vector current form
factor in the semileptonic $\Lambda_b \to \Lambda_c$ decay is
$$\eqalignno{
<\Lambda_c(p') | \bar{c} \gamma_\mu b | \Lambda_b(p)>
  &~=~ \bar{u}(p') [ f_1\gamma_\mu - if_2\sigma_{\mu\nu}q^\nu
                   + f_3q_\mu ] u(p) \cr
  &{\buildrel M\to\infty \over \longrightarrow}~
  [\bar{u}(v') \gamma_\mu u(v)]~ \zeta(v\cdot v') ~~, &(3) \cr}
$$
where $\bar{u}(p')$ and $u(p)$ represent the spinor wavefunctions of the
spin-half baryons. The divergence of this expression yields
$$\eqalignno{
{m_b - m_c \over M_{\Lambda_b} - M_{\Lambda_c}}
  <\Lambda_c(p') | \bar{c} 1 b | \Lambda_b(p)>
  &~=~ \bar{u}(p') [ f_1 + {q^2 \over M_{\Lambda_b} - M_{\Lambda_c}}
                     f_3 ] u(p) \cr
  &{\buildrel M\to\infty \over \longrightarrow}~
  [\bar{u}(v') u(v)]~ \zeta(v\cdot v') ~~. &(4) \cr}
$$
Unlike the meson case, there are no kinematic factors here accompanying
the function $\zeta$ on the right hand side. Therefore, the sum rule
analysis gives only a weak constraint $\zeta'(1) \le 0$.

It is worthwhile to note that a degenerate heavy quark mass limit (i.e.
$m_b = m_c$) can be smoothly taken for both Eqs.~(2) and (4) after
cancelling out the mass factors on the left hand side. This limit makes
the heavy quark flavour symmetry exact, and prevents anomalous dimension
factors (e.g. functions that behave as powers of $\alpha_s(m_c) /
\alpha_s(m_b)$ in leading order weak coupling perturbation theory) from
appearing in the RG analysis of the form factors, thus simplifying the
extraction of the Isgur-Wise functions.

Now we put QCD on a Euclidean lattice using staggered fermions$^9$. The
residual chiral symmetries of this implementation protect the quark mass
operator from unwanted renormalisations. The formalism for performing
strong coupling expansions in lattice QCD is well-known. To keep the
matters simple, the expansion in $1/g^2$ is used often in conjunction
with simultaneous expansions in $1/N_c$ and $1/d$. We follow this approach,
i.e. first obtain results in the limit of infinite number of colours and
infinite number of space-time dimensions and then look at the corrections
due to finite $N_c$ and $d$.

For our purpose, it suffices to look at the extreme strong coupling
limit of the theory. It is described by a trivial fixed point of the
RG evolution. This fixed point is reached by carrying to extreme the
procedure of integrating out all the higher energy virtual states of
the theory while lowering the cutoff. Only the lowest state, described
by a perfectly screened delta-function wavefunction in position space,
survives in each quantum number sector. All the radial excitations
orthogonal to the lowest state, corresponding to extended wavefunctions,
disappear. The interactions amongst the surviving states are of course
altered from their weak coupling behaviour to compensate for the
disappearance of the excited states. We evaluate the correlation
functions in position space using the summation of hopping parameter
expansion method$^{10,11}$ to keep this intuitive picture clear.

The form factors of interest are rather trivial to calculate in the
strong coupling limit. The Feynman diagrams are more conveniently
drawn in terms of colour singlet hadron lines rather than the original
quark lines. Lattice artifacts show up in the formulae (e.g. $\sin(p)$
and $\cos(p)$ functions appear in momentum space propagators instead of
$p$ and $p^2$), but they are easy to keep track of. For $N_c\to\infty$,
the $3-$point correlation function between the external heavy hadrons
and the quark mass operator corresponds to a tree level Y-shaped graph.
For $B\to D$ matrix elements, the three mesons meeting at the vertex
are the incoming $B$, the outgoing $D$ and the scalar $\bar{c}b$. The
full correlation function is merely the product of the three meson
propagators, with the sum in position space over all possible locations
of the vertex producing the constraint of momentum conservation.

In the $d\to\infty$ limit, the $3-$point correlation function after
putting the external hadrons on mass-shell is:
$$
G(D(p';t_1\to\infty),\bar{c}b(q;0),\bar{B}(p;t_2\to-\infty)) ~=~
  {C_D e^{-E_B t_1} \over 2\sinh E_D}~{C_B e^{E_B t_2} \over 2\sinh E_B}~
  {2/\kappa \over 1 + 2\kappa_b\kappa_c \sum_\mu \cos(q_\mu)} ~~,
\eqno(5)$$
where $C$'s are the state normalisation constants, $E$'s are the energies
and $\kappa$'s are hopping parameters representing the renormalised quark
masses (in the notation of Ref.~11). Upon amputating the external legs of
the correlation function we get the desired matrix element, which is the
last factor on the right hand side of Eq.~(5). It can be converted to
the continuum notation by identifying various lattice expressions. With
$$
M_B^2 - M_D^2 ~=~ {1\over\kappa} \left( {1\over\kappa_b}
                                      - {1\over\kappa_c} \right) ~~,~~
m_b   - m_c   ~=~ {1\over2}      \left( {1\over\kappa_b}
                                      - {1\over\kappa_c} \right) ~~,~~
M_{sc}^2 ~=~ {1\over\kappa_b\kappa_c} ~~,
\eqno(6)$$
and the scalar meson mass $M_{sc} \approx M_B + M_D$ in the $M\to\infty$
limit, we obtain
$$\eqalignno{
<D(p') | \bar{c} 1 b | \bar{B}(p)> &~=~ {M_B - M_D \over m_b - m_c}~
  {M_B + M_D \over 1 - q^2/M_{sc}^2} \cr
  &{\buildrel M\to\infty \over \longrightarrow}~
  {M_B - M_D \over m_b - m_c}~{(M_B + M_D)^3 \over 2M_BM_D (1+v\cdot v')}~
  \left( 1 + \CO({1\over M}) \right) ~~. &(7) \cr}
$$
Comparing Eq.~(7) and Eq.~(2), and taking the degenerate heavy quark
mass limit (i.e. $M_B=M_D$), we identify
$$
\xi(v\cdot v') ~=~ \left( {2 \over 1+v\cdot v'} \right)^2 ~~.
\eqno(8)$$
Applying the same method to the quark mass form factor for the spin-half
baryons, we obtain
$$
\zeta(v\cdot v') ~=~ \left( {2 \over 1+v\cdot v'} \right) ~~.
\eqno(9)$$
The difference between $\xi$ and $\zeta$ arises entirely due to
kinematic factors inherent in their definitions.

This analysis shows that in the strong coupling limit the functional
form of the Isgur-Wise functions is completely determined by the
$1/(M_{sc}^2 - q^2)$ dependence, or scalar saturation, of the quark
mass form factor. The remaining hadron mass factors just ensure that
the Isgur-Wise functions are properly normalised, $\xi(1)=1=\zeta(1)$.
Indeed the fact that the normalisations turn out to be correct is a
confirmation of the RG invariance of the form factor evaluation.
The singularity at $v\cdot v'=-1$ is as per expectations, but no
branch cut shows up in this leading order result.

It is straightforward to get rid of the $d\to\infty$ limit, and work
directly in $d=4$. The expressions for state normalisation constants,
hadron masses, hopping parameters and correlation functions become
different in terms of the bare parameters appearing in the lattice
action$^{11}$. The phenomenon of scalar saturation, however, is not
altered at all and the results of Eqs.~(8) and (9) remain valid.

Scalar saturation no longer holds once processes suppressed by
powers of $1/N_c$ are taken into account. Such processes are
represented by diagrams containing hadron loops. For example, to
include the most dominant correction from dynamical light quark
loops we have to evaluate the Feynman diagram with a heavy-light
meson loop attached to three hadron legs, two corresponding to the
external states and one corresponding to the quark mass operator.
This process can be looked upon as the scalar $\bar{c}b$ decaying
into two heavy-light mesons which in turn interact with the two
external hadrons. Evaluation of such diagrams is quite involved,
but we can estimate their magnitude by simple dimensional analysis.
The virtual processes contained in the corrections suppressed by
powers of $1/N_c$ are strong interaction hadron vertices. The
characteristic scale for these vertices (or decay widths) is
$\Lambda_{QCD}$, which remains finite as $M\to\infty$. As a result,
the $1/N_c$ corrections to scalar saturation are expected to be
$\CO(\Gamma_{sc}/M_{sc}) = \CO(\Lambda_{QCD}/M)$. Such corrections
do not contribute to the Isgur-Wise functions which are defined
in the $M\to\infty$ limit. We thus argue that Eqs.~(8) and (9)
are correct even when $N_c$ is finite. A particular instance of
$1/N_c$ suppression is the branch cut for $v\cdot v' < -1$, which
we expect to be softer by a factor of $\Lambda_{QCD}/M$ compared
to the pole at $v\cdot v' = -1$.

Now we can complete the task of inferring what form factors at the
weak coupling fixed point of QCD could have evolved to the strong
coupling expresssions obtained above. $\xi$ and $\zeta$ are
dimensionless functions with fixed absolute normalisations at the
no-recoil point, $v$ and $v'$ are not affected by RG evolution,
and the operator involved in the calculation above (i.e. the local
divergence of the partially conserved vector current in the limit
of degenerate heavy quark masses) was carefully chosen to be RG
invariant to avoid anomalous dimension corrections. We conclude that
the results of Eqs.~(8) and (9) are exact even for the weak coupling
fixed point of QCD.

We close with several comments regarding these results:
\item{(a)}{Since the quark mass operator and its RG evolution plays a
crucial role in our analysis, staggered fermions were necessary in the
lattice implementation. Wilson fermions$^{12}$ would not have been of
much use, since in that case the quark mass operator undergoes additive
renormalisation.}
\item{(b)}{The strong coupling limit does not possess all the properties
of the weak coupling fixed point of QCD, since different quantities
follow different RG evolution patterns. The appropriate choice of an RG
invariant quantity is a crucial ingredient in connecting the two limits.
Here the choice of the vector current was essential; the axial current
does not have as nice renormalisation properties on the lattice. The
Isgur-Wise relations amongst the form factors of the vector and the
axial currents, which hold in the weak coupling limit, are not expected
to hold in the strong coupling limit. This is not a disaster. As long
as one of the form factors can be determined by connecting it from the
strong coupling to the weak coupling limit, the rest can be fixed in
the weak coupling limit by the usual continuum manipulations.}
\item{(c)}{Scalar saturation of the quark mass form factor is a simple
consequence of strong coupling and $N_c\to\infty$ limits. Explicit
lattice formulation is not necessary to infer this behaviour. One can
anticipate it just on the basis of perfect screening between quark and
antiquark, removal of all excited states in the process of RG evolution
and suppression of couplings to multi-hadron virtual states.}
\item{(d)}{The series expansions in $1/g^2$ and $1/M$ have non-zero
radii of convergence, although the one in $1/N_c$ does not. This is
sufficient to avoid any problems while interchanging the order of
limits. Anyhow, the procedure followed here is to first calculate
the form factor in the strong coupling limit, then take the degenerate
quark mass limit, then let $M$ become large to extract the Isgur-Wise
functions, and finally worry about $1/N_c$ corrections.}
\item{(e)}{The light quark mass is kept finite and constant throughout.
It is not at all necessary to take the chiral limit for the light
quarks. The fact that the light quark mass remains a constant in the
$M\to\infty$ limit is enough, for instance to justify the replacement
$M_{sc} \approx M_B+M_D$.}
\item{(f)}{In principle, the complete form factors, i.e. without the use
of $M\to\infty$ limit, can be evaluated in the strong coupling limit.
The subleading corrections suppressed by powers of $1/M$, however, are
not universal and not easy to connect from the strong coupling to the
weak coupling limits. For instance, the terms suppressed by $1/N_c$
have to be kept, and the results have to be converted to continuum
language using light hadron masses, hadronic excitation energies and
widths, and so on. It may be possible to keep track of all this in an
elaborate numerical RG evolution scheme involving many low lying states,
such as the one employed by Wilson to solve the Kondo problem.}
\item{(g)}{The leading Isgur-Wise functions (which have $\xi'(1)=-1$
and $\zeta'(1)=-1/2$) do a reasonable job in fitting the experimental
data for semi-leptonic $B-$decays$^{13,14}$. A better check needs
estimates of symmetry breaking corrections (from unequal heavy quark
masses and from terms suppressed by powers of $1/M$) that have to be
added to the functions $\xi$ and $\zeta$ extracted above. With the
leading term already taken care of, these corrections can be found
using phenomenological models without introducing too much uncertainty
in the final results. Quantitative estimates of the corrections together
with more precise experimental data would provide an accurate test of
the results presented here.}

\vglue 0.6cm
\leftline{\twelvebf Acknowledgements}
\vglue 0.4cm
I am grateful to H.R. Krishnamurthy and Aneesh Manohar for helpful
comments.

\vglue 0.6cm
\leftline{\twelvebf References}
\vglue 0.4cm

\itemitem{1.} N. Isgur and M. Wise, \PLB{232} (1989) 113;
              \PLB{237} (1990) 527.
\itemitem{2.} For a recent review of the subject, see M. Neubert,
              SLAC-PUB-6263 (1993), to appear in Physics Reports.
\itemitem{3.} K. Wilson, \RMP{47} (1975) 773.
\itemitem{4.} For a simple physical renormalisation group picture, see
              P. Nozi\`eres, in {\twelveit Proceedings of the 14th Int.
              Conf. on Low Temperature Physics, Vol. 5}, ed. M. Krusius
              and M. Vuorio, (North-Holland, Amsterdam, 1975), p. 339.
\itemitem{5.} K. Wilson, \PRD{10} (1974) 2445.
\itemitem{6.} See for instance, L. Karsten and J. Smit, \NPB{183} (1981) 103.
\itemitem{7.} J. Bjorken, in {\twelveit Proceedings of the 18th SLAC
              Summer Institute of Particle Physics, 1990}, ed. J. Hawthorne,
              (SLAC Report No. 378, Stanford, 1991), p. 167.
\itemitem{8.} N. Isgur and M. Wise, \PRD{43} (1991) 819.
\itemitem{9.} L. Susskind, \PRD{16} (1977) 3031; \hfil\break
              N. Kawamoto and J. Smit, \NPB{192} (1981) 100.
\itemitem{10.}J.-M. Blairon, R. Brout, F. Englert and J. Greensite,
              \NPB{180} (1981) 439; \hfil\break
              N. Kawamoto, \NPB{190} (1981) 617.
\itemitem{11.}O. Martin, \PLB{130} (1983) 411; \hfil\break
              O. Martin and A. Patel, \PLB{174} (1986) 94.
\itemitem{12.}K. Wilson, in {\twelveit New Phenomena in Subnuclear Physics},
              Erice lectures 1975, ed. A. Zichichi, (Plenum, New York, 1977).
\itemitem{13.}H. Albrecht et al. (ARGUS collaboration),
              \ZPC{57} (1993) 533.
\itemitem{14.}G. Crawford et al. (CLEO collaboration),
              CLEO-CONF 93-30 (1993).

\bye